\documentclass[twocolumn,showpacs,preprintnumbers,amsmath,amssymb]{revtex4}
\usepackage{amsmath, amsthm}
\usepackage{dcolumn}%
\usepackage{graphicx}
\begin{document}

\title{How to make a bilayer exciton condensate flow}

%\maketitle

%% Notice placement of commas and superscripts and use of &
%% in the author list

\author{Jung-Jung Su}
\author{A.H. MacDonald}
%\author{Jung-Jung Su$^{1}$ \& A.H. MacDonald$^{1}$}

%\begin{document}

%\maketitle

\affiliation{Department of Physics, The University of Texas at Austin, Austin, TX 78712, USA}

%\begin{affiliations}
%\item Department of Physics, University of Texas at Austin, Austin TX 78712, USA
%\end{affiliations}

\begin{abstract}
Bose condensation is responsible for many of the most spectacular effects in physics 
because it can promote quantum behavior from the microscopic to the macroscopic world.
Bose condensates can be distinguished by the condensing object; 
electron-electron Cooper-pairs are responsible
for superconductivity, Helium atoms for superfluidity, 
and ultracold alkali atoms in vapors for coherent matter waves.
Electron-hole pair (exciton)
condensation has maintained special interest because it has been difficult to realize experimentally, and because exciton phase 
coherence is never\cite{KohnSherrington} perfectly spontaneous.  Although
ideal condensates can support\cite{Moon} an exciton supercurrent, it has not been
clear\cite{Blatt1962} how such a current could be induced or detected, 
or how its experimental manifestation would be altered   
by the phase-fixing exciton creation and annihilation processes which are inevitably present.
In this article we explain how to induce an exciton supercurrent in separately
contacted bilayer condensates, and predict electrical effects 
which enable unambiguous detection.
\end{abstract}

\maketitle

The order parameter of an exciton condensate is 
\begin{equation} 
\Psi(\vec{r})\; = \; | \Psi(\vec{r})| \exp(i\phi(\vec{r})) \; = \; \langle \, \hat{\psi}_{e}^{\dagger}(\vec{r}) \, \hat{\psi}_{h}(\vec{r}) \; \rangle 
\, = \, \rho(h,\vec{r};e,\vec{r}) 
\end{equation}
where $\hat{\psi}^{\dagger}$ and $\hat{\psi}$ are electron creation and annihilation operators, $\phi(\vec{r})$ 
is the condensate phase, the labels $e$ (electron) and $h$ (hole) refer to the states 
between which phase coherence is established (nearly!) spontaneously, and 
$\rho(h,\vec{r},e,\vec{r})$ is the anomalous density matrix.
Microscopic considerations suggest\cite{Keldysh} 
that spontaneous coherence is likely between a conduction band with occupied
states inside a Fermi surface and a valence band with occupied states outside a
nearly\cite{ContiVignaleMacDonald} identical Fermi surface.
Part of the reason that exciton condensation has not been easy to realize is that
sufficiently perfect nesting between conduction and valence bands is unlikely to occur naturally.
The systems of interest here are artificially fabricated bilayers in which the 
electrons and holes are in well separated two-dimensional electron systems(2DESs), either 
semiconductor quantum wells\cite{quantumwell,Cavendish} or graphene layers\cite{GeimMacDonaldPhysicsToday} separated by a 
dielectric barrier.  The dielectric barrier reduces the strength of exciton creation and 
annihilation processes, and gate control of the density in each layer allows 
electron and hole band Fermi surfaces to be tuned to the same area.
Although simple to describe, this quantum engineering is difficult\cite{quantumwell,Cavendish,graphenebilayer} 
to execute successfully.  Exciton condensation has so far been realized\cite{Spielman2000}
only in the quantum Hall regime in which band dispersion is irrelevant allowing 
spontaneous coherence to occur between spatially separated conduction bands,
or spatially separated valence bands, under circumstances that are achieved routinely.  
The considerations explained in this article apply to quantum Hall exciton condensates 
in the Corbino geometry\cite{TiemannEP2DS,TiemannNJP}, in which current flows across the 2DES bulk, 
but not directly to the Hall bar geometry\cite{RossiAHMprl} in which currents flows along the 2DES edge.

In their pioneering work on exciton Bose condensation Blatt {\em et al.}\cite{Blatt1962} 
argued that because an exciton is neutral, condensation can not lead to {\em spectacular electrical effects}.  
Experimental studies of quantum Hall exciton condensates have already made it 
clear\cite{Spielman2000,jpeahmnature} that this pessimistic 
conclusion is not valid.  The key technical capability not anticipated in 1962 is
the possibility of making independent electrical contact\cite{EisensteinContact} to the 
electron and hole parts of the condensate.  In this article we explain how condensation 
leads to a reorganization of the low-energy charged fermion degrees of freedom which 
is responsible for dramatic changes in the transport properties
of separately contacted condensed bilayers. 

\begin{figure}
\includegraphics[width=1\columnwidth]{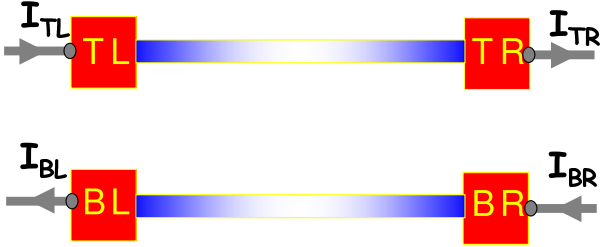}
\caption{\label{fig1}
{\bf Schematic illustration of a separately contacted bilayer exciton condensate}. 
$L$ and $R$ represent the left and right ends of a quantum well system or the 
outer and inner edges of an annular quantum Hall bilayer condensate.
Currents through the four leads can be tuned externally by varying the three independent voltage differences.
Our convention for positive currents is indicated by the arrows attached to the leads.  Transport properties 
are simplest when there is a gap (indicated by white fill in this schematic illustration) in the bulk.}
\end{figure}

Spectacular electrical effects in 
exciton condensates are enabled by the possibility of exciton superflow.  The key issue 
which arises in addressing these phenomena theoretically is understanding how a supercurrent of
neutral excitons can be driven by electrochemical potential differences. 
We consider a bilayer with contacts to both left (L) and right (R) ends of the separate 
quantum wells (top (T=e) and bottom (B=h))  between which coherence is established, as illustrated 
schematically in Fig.[~\ref{fig1}].  (In the case of the Corbino geometry transport of quantum
Hall condensates L and R refer to the outer and inner edges of an annular 2DES.)  For the sake of definiteness we focus our 
attention on voltage biased transport; our analysis is easily extended to the case of current biased 
transport and to systems with additional voltage probes along the sample length.  The observables in
this transport geometry are the currents and voltages in the leads.  Exciton condensation alters 
transport properties by introducing an anomalous symmetry-breaking field in the fermion quasiparticle 
Hamiltonian (more technically in the one-particle Greens function) which enforces inter-layer 
phase coherence, opens up a gap at the Fermi level, 
and allows charge to move freely between layers.
In mean-field-theory, the quasiparticle amplitude for tunneling between top (T) and bottom (B) layers is: 
\begin{equation}
\langle T,\vec{r'}|H_{HF}|B,\vec{r}\rangle = V_{c}(|\vec{r'}-\vec{r}|) \; \rho(T,\vec{r'};B,\vec{r}) 
\end{equation} 
where $V_{c}(r)=e^2/\epsilon\sqrt{r^2+d^2}$ is the inter-layer Coulomb interaction
and $\rho(T,\vec{r'};B,\vec{r})$ is the inter-layer component of the density matrix.
($\rho(T,\vec{r'};B,\vec{r})$ is non-zero only in the condensed state.)
These quasiparticle tunneling 
terms are present even when the microscopic electron Hamiltonian does not include inter-layer transfer processes. 
In a mean-field theory microscopic particle-number conservation in each layer is recovered when the 
exciton pair potential 
is calculated self-consistently; inter-layer quasiparticle currents are then precisely compensated by
exciton condensate supercurrents.  Steady state exciton supercurrents are 
possible, but only when the lead voltages are chosen to satisfy a condition that we explain below. 
Our main predictions are qualitative in nature.  The quantitative properties of a particular exciton condensate system
will depend on the details of its Hamiltonian, including its disorder potentials. We illustrate some of our 
ideas on transport in exciton condensates below by considering a one-dimensional toy model with 
short-range inter-layer interactions, using the NEGF\cite{NEGF} method to evaluate steady-state non-equilibrium density-matrices.

When $\rho(T,\vec{r};B,\vec{r})$ has a spatial phase gradient, the quasiparticle Hamiltonian supports\cite{Moon}
non-zero currents even when the system is in equilibrium, {\em i.e.} when all four chemical potentials are identical.
Because the exciton condensate cannot carry charge, the condensate current (CC) has only\cite{Moon} a counterflow component, {\em i.e.} 
\begin{equation} 
j^{CC}_{T}(x)+ j^{CC}_{B}(x) =0 .
\end{equation}
The total current in each layer can be expressed as the sum of the condensate current and a quasiparticle 
current (QC) driven by lead voltage differences.  The QC conduction is related 
via the Landauer-B\"uttiker formula to the transmission coefficients of quasiparticle waves incident 
from the various leads.  In order to drive steady state condensate current, 
the gate voltages must be chosen so that the counter-flow component of the QC is not spatially constant:
\begin{equation} 
\label{ccons}
[\partial_x (j^{CC}_{T}(x)-j^{CC}_{B}(x))] = -[\partial_x (j^{QC}_{T}(x)-j^{QC}_{B}(x))] 
\end{equation}
Condensate currents are induced by space-dependent quasiparticle counterflow currents. 
This relationship between quasiparticle (QC) and condensate (CC) counterflow currents
follows from the separate conservation of charge in each layer.  Since condensate 
currents cannot enter the leads, they must be present in the bulk when  across the sample$I_{TL}+I_{BL} \ne 0$
or $I_{TR}+I_{BR} \ne 0$.  (Here we use the conventions of Fig.[~\ref{fig1}] for the lead current signs.)
For every electron that is reflected from the top layer to the bottom layer at the left 
edge of the sample, an exciton must be launched.  It follows that $(I_{TL}+I_{BL})$ 
is the exciton supercurrent emitted from the left side of the sample and that $(I_{TR}+I_{BR})$ 
is the exciton supercurrent absorbed on the right.  
Steady state condensate currents are possible only when these two quantities are equal:
\begin{equation} \label{keycondition}
I_{TL}+I_{BL} = I_{TR}+ I_{BR}.
\end{equation} 
In the linear response regime, the four 
lead currents are proportional to the three independent electrochemical potential differences.
Eq.(~\ref{ccons}) places a restriction on the two independent difference ratios.  

\begin{figure}
\includegraphics[width=1\columnwidth]{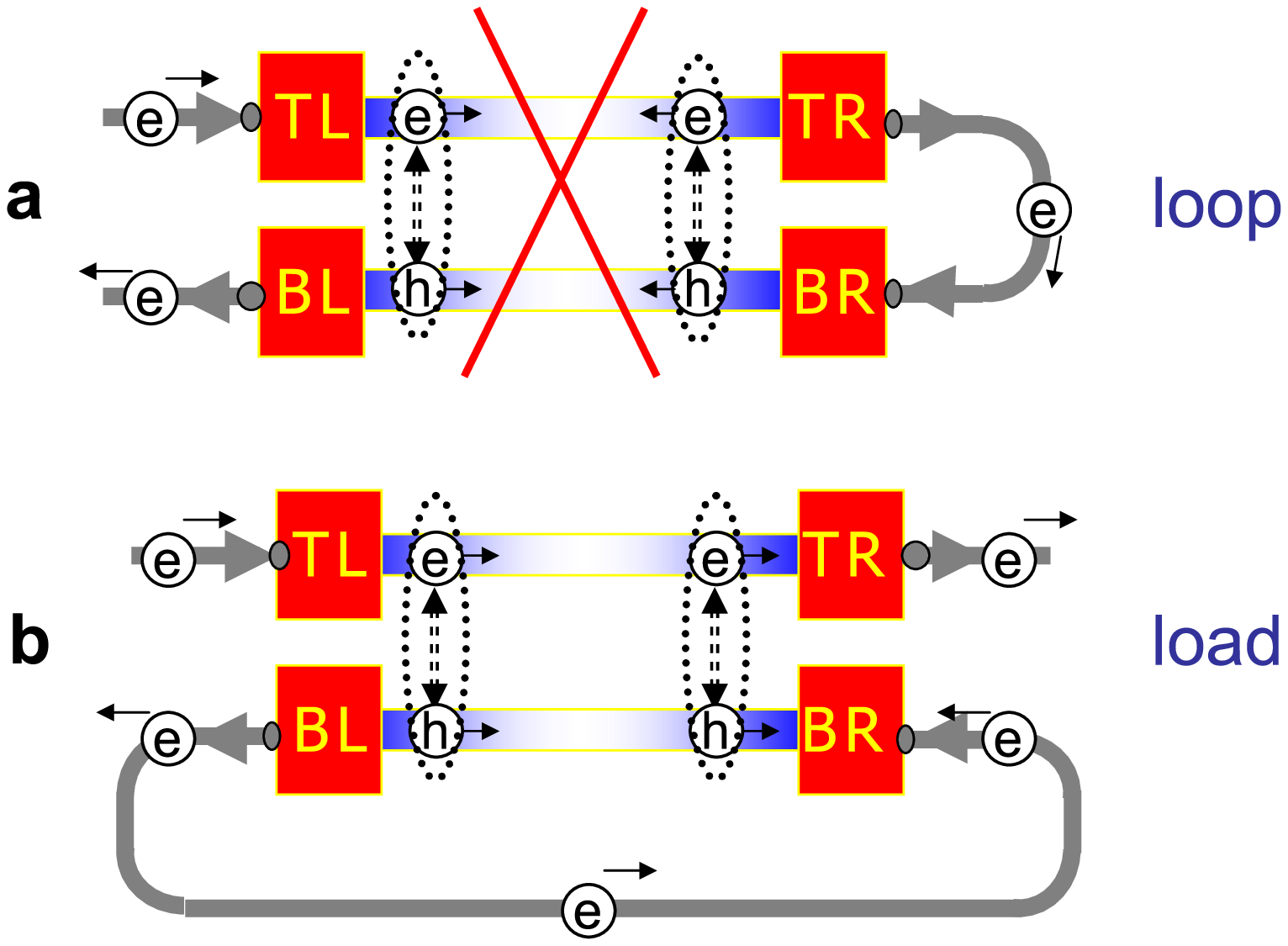}
\caption{\label{fig2}
{\bf Loop and load geometries}. Fig. {\bf a} and {\bf b} 
represent respectively loop and load geometries.
For bilayer exciton condensates only the load geometry {\bf b}
permits steady state exciton superflow.}
\end{figure}

One convenient way to reduce the lead-voltage-space dimension experimentally  
is to connect two contacts with a
resistor $R$, for example in either the loop geometry or the 
load geometry illustrated in Fig.(~\ref{fig2}). 
In the load geometry, the contact voltages are limited to the surface on
which 
$I_{BL} = I_{BR}= (V_{BL}-V_{BR})/R$.  
The consequences of this resistive link between
contacts are easily anticipated in the most common limit in which
the gap due to condensation is large enough or disorder strong enough to prevent quasiparticle 
current conduction across the sample.  Bulk quasiparticle currents are 
expected to be negligible for Corbino geometry transport in the quantum Hall regime, and we anticipate that
they will also be negligible in zero-field bilayer exciton condensates when these are 
ultimately realized.  When quasiparticles cannot flow across the sample, 
$I_{TL}=I_{BL}$ and $I_{TR}=I_{BR}$. 
It follows that Eq.(~\ref{keycondition}) is always satisfied in the load geometry. 
Because it is possible to induce an exciton supercurrent in the bulk of the 
bilayer, a large charge current can flow through the circuit with a resistance due only to the 
load resistor and contact resistances at the two ends of the sample.  In the case of quantum Hall bilayers, we 
predict a Corbino resistance for the load geometry which remains constant as temperature $T \to 0$ 
on bilayer quantum Hall plateaus which are due to exciton condensation\cite{jpeahmnature}.  This 
behavior is in stark contrast with the dramatically increasing resistance expected at low temperatures on plateaus not 
associated with exciton condensation.

Although the loop geometry of Fig.(~\ref{fig2}) is naively  
compatible with steady state counterflow currents, we predict that it cannot support 
steady state exciton superflow because it does not guarantee that the
the exciton supercurrent emitted at the left end of the sample is identical to that absorbed at the right end. 
Indeed for $I_{TR} > 0$ the loop geometry will lead to $V_{TR} > V_{BR}$ 
and therefore to quasiparticle current flowing in the wrong 
direction from top to bottom. 

In the load geometry, the effective two-probe quasiparticle conductances at left and right $G_{L,R} = 
(e^2/h) T_{L,R}$ depend on the details of these contacts and on the quasiparticle Hamiltonian near the sample ends.  
Taking the $TR$ lead as ground, the voltages on the other leads in this circumstance are
$V_{TL}= I \, ( [T_{L}^{-1} + T_{R}^{-1}] (h/e^2) + R )$, $V_{BL}= I \, (R + T_{R}^{-1} (h/e^2))$, and 
$V_{BR} = I \,(h/e^2) T_{R}^{-1}$ where $I_{TL}=I_{BL}=I_{BR}=I_{TR}=I$.
When the load resistor $R$ is small, the current which flows
through the bilayer systems is limited only by the quantum contact resistances $(h/e^2) T_{L,R}^{-1}$.
Even when the microscopic electrons tunneling amplitude is zero, quasiparticles 
move freely between layers allowing charge to be conducted across the highly resistive bulk by 
exploiting the parallel load resistor channel.  

\begin{figure}
\includegraphics[width=1\columnwidth]{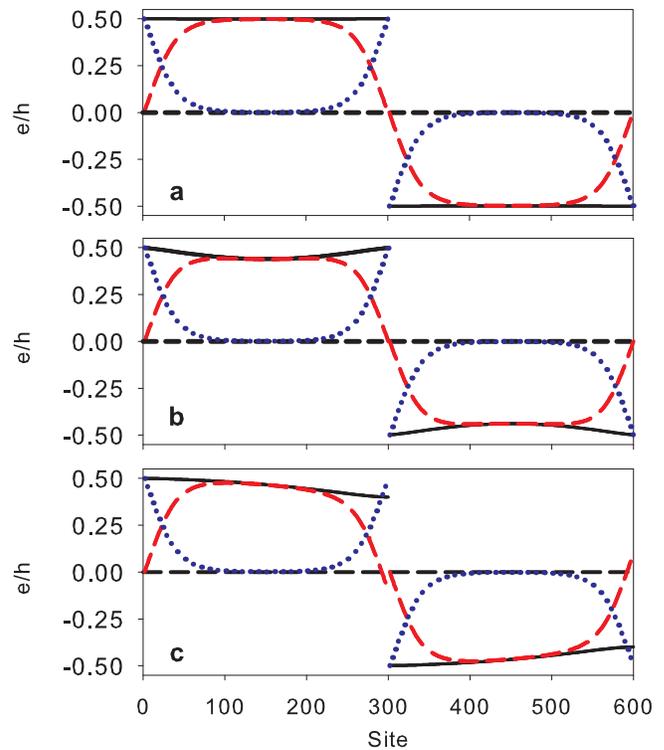}
\caption{\label{fig3}
{\bf 1D Toy model self-consistent current distributions}. 
Sites 1 through 300 label the top layer left to right  
while sites 301 through 600 labels the bottom layer. 
The site separation is $0.08$ Bohr radii.
The black solid lines, red dashed lines, and blue dotted lines represent 
the total current, the condensate current (CC) and the quasiparticle current 
respectively.  In {\bf a},$V_{TL}=1.0, V_{TR}=-0.5, V_{BL} = V_{BR}=0.0$ and the 
bare interlayer tunneling amplitude $\Delta_t=0.0$.
(Voltages are in ${\rm Rydberg}/e$)  The total (CC+QC) current is 
constant and the QP current is very small near the system center.
{\bf b},$V_{TL}=0.5, V_{TR}=-0.5, V_{BL}=0.0, V_{BR}=0.0, \Delta_t=0.005$.
% Final:  Need to be consistent between \Delta_t and t? 
When bare tunneling occurs the total current is no longer a constant, but 
the transport characteristics in loop geometry do not change in a qualitative way.
{\bf c},$V_{TL}=0.5, V_{TR}=-0.5, V_{BL}=0.0, V_{BR}=-0.1, \Delta_t=0.005$.
Steady state solutions with net exciton creation rates are possible 
because of bare electron interlayer tunneling.
The unit of voltage is ${\rm Rydberg}/e$. 
}
\end{figure}  

In Fig.(~\ref{fig3}) we plot numerical results for the self-consistently calculated 
quasiparticle and condensate current distributions in a one-dimensional tight-binding-model  
bilayer system with an electron band in the top layer and an equal density hole band in the bottom layer.
The inter-layer tunneling amplitude in the quasiparticle Hamiltonian of this model is 
\begin{equation} 
\langle T,I'|H_{HF}|B,I\rangle = \delta_{I',I} \, V \,  \rho(T,I';B,I) 
\label{meanfieldtheory}
\end{equation}  
where $V$ is the short-range inter-layer interaction strength and 
$I,I'$ is the site indices.  Because of the energy gap induced
in the quasiparticle Hamiltonian by this tunneling term, the quasiparticle current is deflected to flow between
T and B leads at both left and right.  The numerical results in Fig.(~\ref{fig3}) have been obtained by 
using mean-field theory (Eq.(~\ref{meanfieldtheory})) to calculate the quasiparticle Hamiltonian given
the density matrix, and the NEGF formalism\cite{NEGF} to evaluate the non-equilibrium density matrices given the lead voltages and the
quasiparticle Hamiltonian.  In the top panel of Fig.(~\ref{fig3})
%we have chosen $V_{TR}$ as ground, and
% Jung-jung: I chose the zero of V so that it is consistent with the caption.  
,$V_{B(L,R)} = ((V_{TL}+V_{TR}) \pm V')/2$.  Because the left and right sides of our model system 
are identical this choice produces $I_{BL}=I_{BR}$, corresponding to the load 
geometry.  ($V'$ is related to the load wire resistance by
 % $V'/V_{TL} = R/(R+ T_{L}^{-1} + T_{R}^{-1})$.)  
  $V'/(V_{TL}-V_{TR}) = R/(R+ T_{L}^{-1} + T_{R}^{-1})$.)  
When the non-equilibrium mean-field calculation is carried to self-consistency, 
the phase of the order parameter develops a spatial gradient and the system
carries a steady state condensate current.  

The total current is constant in each layer as 
required by charge conservation.  In the ordered state, the low-energy charged degrees of 
freedom are reorganized from bare electrons which cannot transfer between layers to exciton-condensate 
quasiparticles which occupy the two-layers simultaneously and coherently.  This 
profound reorganization of low-lying charged excitations 
decreases the resistance exponentially compared to the case in which the gap in the quasiparticle 
spectrum is not due to exciton condensation.  In the quantum Hall case for example, we predict
load geometry resistances that are orders of magnitude smaller at $\nu=1$ (the exciton condensation case) than
at $\nu=2$.  For the loop geometry, no self-consistent steady-state solutions of the non-equilibrium 
self-consistent field equations exist.  In this case we anticipate enhanced noise due to order-parameter 
time dependence and dramatically higher steady state time-averaged resistance. 

As emphasized in a fundamental early paper by Kohn and Sherrington\cite{KohnSherrington}, 
phase coherence between different bands in a solid can never be completely spontaneous.  
In the case of bilayer condensates, inter-layer electron tunneling, which creates or 
annihilates excitons, is expected to be the dominant process which fixes a preferred phase 
and violates separate charge conservation in each layer.  As illustrated in panel {\bf b} 
in Fig.(~\ref{fig3}), however, weak inter-layer electron tunneling has little effect on
transport properties.  Most experimental anomalies associated with
bilayer exciton condensation require only that the quasiparticle 
tunneling amplitude be dramatically enhanced compared to its bare values.  
As we now explain the main consequence of bare-electron inter-layer tunneling is partial relaxation of the 
current-conservation condition, Eq.(~\ref{keycondition}).

The role of a bare electron inter-layer tunneling amplitude is most simply discussed 
using the minimal field theory model\cite{Moon} of a bilayer exciton condensate: 
\begin{eqnarray}\label{EnSG}
E= \int dr \ \left[
\frac{\hbar^2 \rho_s}{2 m^*} |\nabla \theta|^2
  - M \, \Delta_t \cos \theta \
\right]
\end{eqnarray}
where $\theta$ is the inter-layer coherence phase, $\Delta_t$ is the bare-electron tunneling amplitude, 
,$m^*$ is the effective mass in the Landau-Ginzburg type theory,
$M = 2 |\rho(T,B)|$ is the condensate order parameter, and 
$\rho_s$ is its superfluid density.  In this description 
$j=(\hbar \rho_s \partial_x \theta) /m^*$ is the exciton condensate supercurrent. 
Minimizing this energy functional leads to a sine-Gordon equation from which it follows that 
\begin{eqnarray}
\partial_x \left[ 
\frac{1}{2} \lambda^2 (\partial_x \theta)^2
+ \cos \theta \right]=0.
\label{COM}
\end{eqnarray}
where $\lambda^2 = (\rho_s/M) (\hbar^2 /{ m^* \Delta_t})$ is the model's 
Josephson length.  From the constant-of-motion in square brackets in Eq.(~\ref{COM}), 
we obtain an explicitly expression for the difference between the condensates 
currents at opposite ends of the sample: 
\begin{eqnarray}
j_R^2- j_L^2 =  (\cos\theta_L-\cos \theta_R)
\ \frac{2}{\lambda^2} \ \left( \frac{\hbar \rho_s}{m^*} \right)^2
\end{eqnarray}
Defining $J \equiv (j_R+j_L)/2$ and $ \delta j \equiv j_R-j_L$,
we conclude that steady state collective currents are 
possible provided that the exciton currents injected at left and 
absorbed at right differ by less than  
\begin{eqnarray}
\delta j_{\rm max}  &\le& \Delta_t \; 
\frac{2 \rho_s M}{m^* } \ \frac{1}{J} \; .
\label{jmax}
\end{eqnarray}
The limit on $\delta j$ is closely analogous to the critical current
predicted\cite{RossiAHMprl} for inter-layer tunneling.
For contact resistances are $\sim h/e^2$ we predict that
large conductances will occur when $V_{BR}$ is varied 
by less than $ \sim (h/e) \, \delta j_{max}$ from its load resistance value.
In our one-dimensional toy model for example ($j_{L}=I_{TL}/e$, $j_{R}=I_{TR}/e$),
Eq.(~\ref{jmax}) implies that $\delta j_{\rm max} \sim 0.1({\rm Rydberg}/h)$
% Final:  Is this all with m^* equal electron mass.  If so maybe we should 
% just say use m instead of m^*?  and say that it is the electron mass? 
% Jung-Jung: 
% 1. m^* is the 'effective mass' in the macroscopic theory
%    in Tinkham it is set to be twice the free electron mass
% 2. I have corrected the unit of j to be (Ryd/h) hereafter

for $\Delta_t=0.005 \ ({\rm Rydberg})$ and total current $j=0.5 ({\rm Rydberg} /h)$.
Indeed, our NEGF calculations find a self-consistent solution, illustrated in Fig.(~\ref{fig3} {\bf c}),
for the voltage configuration $V_{TL}=0.5,V_{BL}=0.0,V_{TR}=-0.5,V_{BR}=-0.1$ 
(in (${\rm Rydberg}/e$) units) (corresponding to $\delta j \sim 0.1 ({\rm Rydberg}/h)$) 
but not for larger $\delta j$ in qualitative agreement with this prediction.

This work has been supported by the Welch Foundation and by the National Science Foundation under grant DMR-0606489.
   AHM acknowledges long-standing stimulating interactions with Jim Eisenstein, Werner Dietche,
   and Klaus von Klitzing which have informed this analysis and helpful discussions with Mike Lilly, Lars Tiemann, and 
Ian Spielman.

\vspace{0.6cm}
\noindent

%% Put the bibliography here, most people will use BiBTeX in
%% which case the environment below should be replaced with
%% the \bibliography{} command.

%\bibliography{Biblio.bib}
%\bibliography{../drafts/Biblio.bib}

\begin{thebibliography}{99}
\bibitem{graphenebilayer} The increase in electrostatic potential
between electron and hole layers required to engineer an exciton condensate is qualitatively smaller for
graphene layers because they are gapless. 
At least in this sense, the requirements for engineering exciton condensation are less severe
for graphene layers then for semiconductor quantum wells.
\end{thebibliography}

%\begin{thebibliography}{1}
%\bibitem{dummy} Articles are restricted to 50 references, Letters
%to 30.
%\bibitem{dummyb} No compound references -- only one source per
%reference.
%\end{thebibliography}

%% Here is the endmatter stuff: Supplementary Info, etc.
%% Use \item's to separate, default label is "Acknowledgements"

%\begin{addendum}
% \item[Competing Interests] The authors declare that they have no
%competing financial interests.
% \item[Correspondence] 
%should be addressed to J.J.S.~(email: jung@physics.utexas.edu).
%\end{addendum}

%%
%% TABLES
%%
%% If there are any tables, put them here.
%\begin{table}%
%
%\end{table}

\end{document}